\documentstyle[preprint,aps]{revtex}                  %%%To be commented
\def\ltap{\ \raisebox{-.5ex}{\rlap{$\sim$}} \raisebox{.4ex}{$<$}\ }
\def\gtap{\ \raisebox{-.5ex}{\rlap{$\sim$}} \raisebox{.4ex}{$>$}\ }
\begin{document}
\pagestyle{empty}                                      %%%To be commented
\preprint{
\font\fortssbx=cmssbx10 scaled \magstep2
\hbox to \hsize{
%\hbox{\hskip.2cm  hep-ph/9610xxx}
\hfill%$\raise 1cm\vtop{\hbox{CERN-TH/95--326}
            %    \hbox{FISIST/14-95/CFIF}
                %\hbox{NTUTH-95-11}\hbox{}}$}
\hbox{NTUTH-96-11}}
}
\draft
\vfill
\title{
Glueballs: Charmonium Decay and $\bar p p$ Annihilation
}
\vfill
\author{Wei-Shu Hou}
\address{
Department of Physics, National Taiwan University,
Taipei, Taiwan 10764, R.O.C.
}
\date{\today}
%
%\vskip -1cm
%
\vfill
\maketitle
\begin{abstract}
The vector glueball $O$, made of 3 valence gluons,
is expected to be  ``clean":
it mixes less with quarkonia, but mediates OZI violations.
The recent $0^{++} $ glueball candidate and
the persistence of the $J/\psi,\ \psi^\prime \to \rho\pi$ puzzle
suggest $m_O\simeq m_{J/\psi}$, with mixing angle $\sim 2^\circ$--$ 4^\circ$,
hence $\Gamma(O\to \rho\pi$, $K^+K^-$, $e^+e^-) \sim$
MeV, few keV, few eV.
Lower and upper bounds on $\Gamma_O$
can be argued from $e^+e^- \to \rho\pi$ energy scan data and the
condition $B(O\to \rho\pi) > B(J/\psi\to \rho\pi)$.
$O$ dominance may explain the ``large" OZI violation in
$^1S_0(\bar pp)\to \phi\gamma$ vs. $\omega\gamma$.
\end{abstract}
%
%\vfill
%
%
%\narrowtext
%
\pagestyle{plain}

Glueballs are fundamental objects in the sense that, if all quarks were
as heavy as charm or bottom, we would still have neutral, quarkless mesons
starting around 1--2 GeV in mass, and the lowest lying ones would be stable!
Our world is complicated, however, by the existence of an approximate 
flavor $SU(3)$ symmetry at the QCD scale.
The abundance of $q\bar q$ meson states in the 1-2 GeV region
and glueball--quarkonium mixings make the identification of the
would-be lightest neutral hadrons extremely difficult.
To date, we have not yet established any glueball state beyond doubt.

There has been, however, some recent progress \cite{Landua}
in the $0^{++}$ scalar glueball sector,
where experiment and lattice results are converging.
On one hand, in part due to high statistics studies of 
$\bar pp \to \pi^0 M\bar M$ modes \cite{Landua},
there is now an excess of isoscalar
$0^{++}$ mesons, namely, 
$f_0(1370)$, $f_0(1500)$ and $f_0(1720)$ \cite{Zhu}.
Together with the $I= 1/2$ and $1$ mesons $K_0^*(1430)$ and $a_0(1450)$,
they do not all fit into a $q\bar q$ nonet \cite{Landua}.
On the other hand, recent lattice calculations predict \cite{Landua}
the $0^{++}$ glueball mass to be 
$1600 \pm 100$ MeV.
Although two groups \cite{UKQCD,Weingarten}
claim opposite ends of the above range,
their close agreement is in fact quite remarkable.
There are thus competing claims that either \cite{AC} $f_0(1500)$
or \cite{SVW} $f_0(1720)$ is the $0^{++}$ scalar glueball $G$
while the other is dominantly $s\bar s$.
It is likely, however, that
both states have large glueball admixtures \cite{Landua}.
As to the lattice expectation \cite{Landua} of
$2200 - 2500$ MeV for a $2^{++}$ glueball,
further evidence for the $\xi(2230)$ state has
been reported \cite{Landua,xi} recently.
All these states are seen in $J/\psi \to \gamma + X$ transitions \cite{Landua},
where the ``glue content" \cite{Farrar} appears to be high.

The  $0^{++}$ and  $2^{++}$ are 2-$g$ glueballs in the constituent picture.
They are clearly difficult to disentangle from nearby quarkonia.
In this paper we are mainly concerned with
the lowest lying $1^{--}$ glueball state called $O$,
which can only be made of 3 constituent gluons.
Because of its composition, and because it should be heavier,
we expect it to mix less with $q\bar q$ mesons.
It should therefore retain more its glueball character,
hence cleaner and easier to interpret \cite{HS}
once it is seen.
Unfortunately, such glueballs are harder to produce since they 
require 3 gluons to construct.
This brings us naturally, however, to vector charmonium decay,
which, according to perturbative QCD, proceeds via 3 gluons.
Interestingly, there has long been \cite{Franklin}
some ``anomaly" in $J/\psi$ vs. $\psi^\prime$
decays that seem to call for the existence of $O$.
Assuming that $J/\psi$, $\psi^\prime\to 3g \to X$ differ
only in the $c\bar c$ wave function at the origin,
the ratio of branching ratios is expected to follow the so-called
$15\%$ rule,
\begin{equation}
R_{\psi^\prime\psi} \equiv B(\psi^\prime\to X)/B(J/\psi\to X)
                     \simeq B(\psi^\prime\to e^+ e^-)/B(J/\psi\to e^+ e^-)
                     \cong  0.15,
\label{rule}
\end{equation}
which holds for $p\bar p$, $p\bar p + n\pi$,
$5\pi$, $7\pi$, and the recently reported $b_1\pi$ \cite{AP} and $\phi f_0$ \cite{Gu1} modes.
However, as originally reported by MARK II \cite{Franklin},
and reconfirmed by BES,
although quite abundant in $J/\psi$ decay ($\sim 1\%$),
the $VP$ modes $\rho\pi$ and $K^*\bar K$
are not yet seen for $\psi^\prime$ \cite{Gu1}
\begin{equation}
B(\psi^\prime \to \rho\pi) < 2.9 \times 10^{-5},\ \ \ \ 
B(\psi^\prime \to K^{*+}K^-) < 3.2 \times 10^{-5}.
\label{psip}
\end{equation}
A similar situation now seems \cite{Gu1} to be emerging for $VT$ modes
such as $\omega f_2$, $\rho a_2$ and $K^*\bar K_2$.
The simplest and most attractive explanation is \cite{HS} to invoke a nearby
3-$g$ resonance $O$ that (see Fig. 1)
enhances greatly the $J/\psi$ decay into these anomalous channels.
However, the BES experiment has recently reported \cite{scan} an 
energy scan of $J/\psi \to \rho\pi$, which appears to rule out
the vector glueball $O$ in the so-called Brodsky-Lepage-Tuan (BLT)
domain \cite{BLT}
\begin{equation}
\vert m_O - m_{J/\psi} \vert < 80\ \mbox{MeV},\ \ \ \
\Gamma_O < 160\ \mbox{MeV}.
\label{BLT}
\end{equation}
In this paper, we make a careful assessment of these recent data.
We find that $0^{++}$ data and $\psi^\prime$ results
support $m_O \simeq m_{J/\psi}$,
while the conclusion drawn from the BES scan is questionable.
A consistent decay scenario for $O$ emerges.
Further evidence for $O$ is argued from the so-called
$\bar p p \to \phi + X$ vs. $\omega + X$ anomaly.

Shortly after the $J/\psi$ discovery,
Freund and Nambu (FN) postulated \cite{FN} the existence of a state $O$ 
which mediates the Okubo-Zweig-Iizuka (OZI) \cite{OZI} violating
$\phi \to \rho\pi$ decay (assuming ideal $\phi$--$\omega$ mixing).
This is a ``pomeron daughter",
a ``closed string without quarks",
hence a ($1^{--}$) glueball in present terms.
Its mass was argued from dual dynamics to be $ \sim 1.4 - 1.8$ GeV,
and $J/\psi \to \rho\pi$ was predicted to be a dominant decay mode.
From a constituent gluon picture, the low lying 3-$g$ glueball spectrum
was studied \cite{HS} by Hou and Soni (HS),
assuming two-body forces only.
%,
%with the $g$-$g$ potential extracted in the usual way.
Taking the constituent mass
$m_g \sim 500$ MeV \cite{Cornwall}, it was found that
$m_O \cong 4.8\, m_g \simeq 2.4$ GeV,
which is considerably heavier than the estimate of FN.
As $\Gamma(J/\psi\to \rho\pi) \cong 1.1$ keV
turned out to be much smaller than predicted,
 the $O$-$V$ mixings (Fig. 1) were allowed to have
QCD motivated scale dependence \cite{HS}
\begin{equation}
f_{O\omega} : f_{O\phi} : f_{O\psi} = 
(\sqrt{2} : -1 : 1)f(q^2).
\label{f}
\end{equation}
To explain the freshly reported \cite{Franklin} $\rho\pi,\ K^* \bar K$ anomaly,
HS invoked \cite{HS} a pole dominance, or resonance enhancement model:
 %\begin{enumerate}
 %\item 
i) $J/\psi \to O \to \rho\pi \gg J/\psi \to ggg\vert_{\rm cont.} \to \rho\pi$;
 %\item 
ii) $J/\psi \to O \to \rm{other} \ll J/\psi \to ggg\vert_{\rm cont.} \to other$;
 %\item 
iii) $\psi^\prime \to O \to \rm{ANY} \ll \psi^\prime \to ggg\vert_{\rm cont.} \to \rm{ANY}$,
 %\end{enumerate}
where ``cont." stands for continuum,
and likewise for $K^*\bar K$.
This leads to the ratio
\begin{equation}
{\Gamma(\psi^\prime\to O \to \rho\pi) \over \Gamma(J/\psi\to  O \to\rho\pi)}
\cong {\left(m_{\psi}^2 - m_O^2 \over
m_{\psi^\prime}^2-m_O^2\right)^2} 
{f_{O\psi^\prime}^2\over f_{O\psi}^2}.
\label{ratio}
\end{equation}
As the anomaly deepened, implying \cite{comm2} that $O$ has to be
rather degenerate with $J/\psi$, BLT \cite{BLT}
included the $O$ width $(m_V^2 - m_O^2)^2
 \longrightarrow (m_V^2 - m_O^2)^2 + m_O^2\Gamma_O^2/4$
and argued that
the range of eq. (\ref{BLT}) was implied.
Fortuitous as it may seem, 
recent BES data \cite{Gu1} on $\psi^\prime \to VP$ modes, eq. (\ref{psip}),
continue to support this.
We now wish to argue from $0^{++}$ data that the range of eq. (\ref{BLT})
is also motivated from outside of charmonium physics.

One of the main uncertainties in the potential model
is the constituent, or dynamical mass $m_g = 500 \pm 200$ MeV \cite{Cornwall}.
The $0^{++}$ glueball $G$ is predicted \cite{CS} to have mass
$m_G \simeq 2.3\, m_g$ while the $1^{--}$ glueball $O$ has mass \cite{HS}
$m_O \simeq 4.8\, m_g$.
Perhaps the ratio $m_O/m_G \simeq 2.1$ is more trustworthy.
Taking $m_G = 1400,\ 1500,\ 1600$ MeV,
we find $m_g \simeq 610,\ 650,\ 700$ MeV respectively, which is remarkably
close to twice the constituent quark mass $2m_q$.
The predicted $O$ mass then is
$2920,\ 3130,\ 3340$ MeV, respectively,
which is right in the ballpark \cite{mO} of eq. (\ref{BLT})!
The consequences of this upward shift from the original
HS paper to eq. (\ref{BLT}) turns out to be more self-consistent
and in better agreement with data,
but were not explored in detail by BLT.
First, HS advocated \cite{HS} the direct search via $J/\psi,\ \psi^\prime \to G+ O$.
Now that $m_G + m_O > m_{\psi^\prime}$,
these modes are clearly forbidden,
although $\psi^\prime \to (\pi\pi)_{I=0} + O\ (\to \rho\pi,\ K^*\bar K)$
search should continue.
Perhaps one can search for $\Upsilon(1S) \to G + O\ (\to \rho\pi,\ K^*\bar K)$
at CLEO and at future B Factories.
Second, assuming $O$ saturation, the known 
$J/\psi\to \rho\pi$ width provides an important constraint,
 %\begin{equation}
$f(m_{J/\psi}^2) \cong 0.025\ \rm{GeV}\times
((m_{J/\psi}^2 - m_O^2)^2 + m_O^2\Gamma_O^2/4)^{1/4}$,
 %\end{equation}
which leads to the mixing angle
\begin{eqnarray}
\sin\theta_{O\psi} &\simeq& 
f(m_{J/\psi}^2)/\sqrt{(m_{J/\psi}^2 - m_O^2)^2 + m_O^2\Gamma_O^2/4}  \nonumber \\
&\simeq& 0.025\ \rm{GeV}/\left((m_{J/\psi}^2 - m_O^2)^2
                                                       + m_O^2\Gamma_O^2/4\right)^{1/4}.
\end{eqnarray}
%which is quite small.
%For example,
Refining the mass range of eq. (\ref{BLT}) to
(the bounds on $\Gamma_O$ would be explained later)
\begin{equation}
20\ \mbox{MeV} < \vert m_O - m_{J/\psi} \vert < 80\ \mbox{MeV},\ \ \ \ 
4\ \mbox{MeV}\ltap \Gamma_O \ltap 30 - 50\ \mbox{MeV},
\label{Orange}
\end{equation}
since the degeneracy of $m_O$ to within
20 MeV of $J/\psi$ would be too fortuitous, we find
\begin{equation}
0.035 < \sin\theta_{O\psi} < 0.071,
\end{equation}
which is reasonably small. 
In what follows, we shall use $m_O = 3180$ MeV for numerical illustration,
where $f(m_{J/\psi}^2) \simeq 0.018$ GeV$^2$
and $\sin\theta_{O\psi}\simeq 0.034$.

One can now see that, because of the paucity of isocalar 
$1^{--}$ mesons,  $J/\psi$-$O$ mixing introduces
the chief $q\bar q$ content to the state $O$,
while $\psi^\prime$, $\phi$, $\omega$ mixings with $O$ are suppressed by
propagator factors.
The two physical states $O$ and $J/\psi$ can be written as \cite{AGK,CGT}
\begin{eqnarray}
\vert J/\psi\rangle & \cong &   + \cos\theta_{O\psi}\; \vert c\bar c(1S)\rangle
                                                       + \sin\theta_{O\psi}\; \vert ggg\rangle, \nonumber  \\
\vert O\rangle \ \ \;  & \cong & - \sin\theta_{O\psi}\; \vert c\bar c(1S)\rangle
                                                      + \cos\theta_{O\psi}\; \vert ggg\rangle,
\label{Opsi}
\end{eqnarray}
where $\vert c\bar c\rangle$ and $\vert ggg\rangle$ are
pure $c\bar c$ and $ggg$ states.
The pole dominance model with near degeneracy of $O$ and $J/\psi$
then implies that
$\Gamma(O\to \rho\pi) \simeq
 \Gamma(J/\psi \to\rho\pi)/\sin^2\theta_{O\psi} \simeq 1$ MeV,
and similarly $\Gamma(O\to K^*\bar K) \simeq 0.7$ MeV.
In contrast, $O\to e^+e^-$ proceeds via its $c\bar c$ content,
hence $\Gamma(O\to e^+e^-) \simeq
 \Gamma(J/\psi \to e^+e^-)\times\sin^2\theta_{O\psi} \simeq 6$ eV,
which is extremely small.
This is in strong contrast to usual neutral $q\bar q$ mesons \cite{HS,FN}.
Assuming $\omega$ \cite{HS} and  $\phi$ \cite{FN} dominance, respectively,
one finds that $\Gamma(O\to p\bar p) \simeq 10$ keV
and $\Gamma(O\to K\bar K) \sim 6$ keV,
which is much smaller than $\rho\pi$ and $K^*\bar K$ modes.
These numbers fit the resonance enhancement model
prescription for the $\rho\pi$ anomaly fairly well.

A generic lower bound on $B(O\to \rho\pi)$,
hence an upper bound on $\Gamma_O$, can be argued
from $J/\psi$ and $\psi^\prime$ data.
If $B(O\to \rho\pi) \ltap B(J/\psi\to \rho\pi) \sim 1\%$,
then the $\vert ggg\rangle$ component of $J/\psi$
would saturate the $J/\psi$ width,
and many more modes would violate the 15\% rule of eq. (\ref{rule}).
Since this is not the case, we expect $B(O\to \rho\pi) \gtap $ few \%,
and $\Gamma_O \ltap 30$ MeV
(hence the upper bound on $\Gamma_O$ in eq. (\ref{Orange}))
for $m_O = 3180$ MeV,
which is relatively narrow for such a heavy flavorless hadron.
The bound on $\Gamma_O$ decreases as $m_O\to m_{J/\psi}$.

The immediate question to address is the
absence of evidence for an $O$ state in the vicinity of $J/\psi$. 
Scanning the $J/\psi\to \rho\pi$ mode over a 40 MeV energy interval,
the  BES experiment has recently reported \cite{scan} the bound 
\begin{equation}
\sigma_{O+I}/\sigma_{J/\psi} < 0.098,
\label{scan}
\end{equation}
at the 90\% confidence level,
where $\sigma_{O+I}$ is the extra cross section
due to $O$ and its interference with $J/\psi$
in the energy window.
After some analysis, BES claims \cite{scan} that a
broad and nondegenerate $O$ (with $J/\psi$) is ruled out.
This is quite puzzling, since intuitively
a broad state not too close to $J/\psi$ should
have been harder to discern.
Note that, according to eq. (\ref{Opsi}),
$e^+e^- \to J/\psi \to \rho\pi$ and $e^+e^- \to O \to \rho\pi$
should have equal total cross sections (see Fig. 2),
but the peak cross section for the latter is far less than the former,
weighed down by the factor $\Gamma_{J/\psi}^2/\Gamma_O^2$.
This is born out by our numerical example.
On closer inspection, one finds that the assumption stated in eq. (8) of ref. \cite{scan},
viz. $B(\psi(2S) \to \rho\pi)/(\sigma_{J/\psi}/\sigma_{\rm tot}) \simeq 0.15$
is self-contradictory, since it ignores the $\vert ggg \rangle$ content
of the physical state $\vert J/\psi\rangle$
which is responsible for $J/\psi\to \rho\pi$ enhancement.
Taking eq. (\ref{Opsi}) into due account,
the scan result of eq. (\ref{scan}) cannot rule out 
a glueball state $O$ with $\vert m_O - m_{J/\psi}\vert$ and $\Gamma_O$
greater than a few times the BES energy resolution
$\Delta E \simeq 2$ MeV \cite{radtail,update}.
Preliminary analysis along similar lines in the search for $O$
in the $\rho\pi$ invariant mass spectrum
of $\psi^\prime \to \pi^+\pi^- + \rho\pi$ decay leads to the bound \cite{Olsen}
$\Gamma_O > 4$ MeV for $m_O \simeq 3180$ MeV, implying that
$B(O\to \rho\pi) < 25\%$.

Collecting results, we find $4$ MeV $\ltap \Gamma_O < 30-50$ MeV,
and  few $\% \ltap B(O\to \rho\pi) \ltap 25\%$ for $m_O \simeq 3180$ MeV.
Unlike the old \cite{HS} result of HS, where $\Gamma(O\to \rho\pi)$
was estimated to be $\sim 50$ MeV
(since $m_O$ was far away from $m_{J/\psi}$),
hence must be a predominant decay mode,
our present result of a dominant but not predominant
$O\to \rho\pi$ mode is more plausible.
The smallness of $\Gamma_O$ is in part because
$O\to GG$ is phase space and $P$-wave suppressed.

But what about the emerging $VT$ anomaly \cite{Gu1},
where the $\omega f_2$, $\rho a_2$ and $K^*\bar K_2$
modes are also seen to be suppressed in $\psi^\prime$ decays?
Note that the observed $J/\psi \to VP,\ VT$ and
$\eta_c \to VV$ modes are all rather prominent,
each of order $1\%$.
As suggested by Anselmino, Genovese and Kharzeev \cite{AGK},
the $\eta_c$ could also mix with a $0^{-+}$ 3-$g$ glueball
(containing sizable 2-$g$ content \cite{HS}, which explains the
large $\eta_c \to VV$ width compared to $J/\psi \to VP$),
which mediates the $VV$ modes.
Interestingly, the potential model predicts \cite{HS} altogether
{\it four} lowest lying 3-$g$ glueballs:
one pseudoscalar $0^{-+}(1)$,
{\it two} vectors $1^{--}(0)$ and $1^{--}(2)$,
and a spin 3 state $3^{--}(2)$,
where the number in parenthesis is
the total spin of any $(gg)_8$ pair.
They are all roughly {\it degenerate} at
$4.8\, m_g \simeq m_{J/\psi}\simeq m_{\eta_c}$.
It appears then that the two final state mesons tend
to ``remember" the original spin configurations.
One might picture the glueball as decaying
via $g_8(gg)_8 \to (q\bar q)_8 (q^\prime\bar q^\prime)_8$,
and the $q$ and $q^\prime$ undergoes
some Fierz-like rearrangement before hadronizing.
With the near degeneracy of the $1^{--}$ and
$0^{-+}$ 3-$g$ states to $J/\psi$ and $\eta_c$,
one has a crude but common scenario
for the observed prominence of $J/\psi\to VP,\ VT$
and $\eta_c\to VV$ modes.
The $VP$ and $VV$ modes are probably highly suppressed in
$\psi^\prime$ and $\eta_c^\prime$ decays by
total hadron helicity conservation (HHC) \cite{BL}.
The observation \cite{Gu1} of a suppressed but nonvanishing 
$\psi^\prime \to \rho a_2$ mode is consistent with the
$\psi^\prime \to VT$ modes being allowed by HHC,
and the absence of a nearby $O$-pole \cite{comm2}.

Recall that $O$ was originally introduced \cite{FN}
to explain OZI  \cite{OZI} dynamics.
Indeed, using eq. (\ref{f}) one obtains from $\phi\to \rho\pi$ decay
$f(m_\phi^2) \sim 0.5$ GeV$^2$,
which is of typical hadronic scale,
but $f(m_\phi^2)/(m_O^2 - m_\phi^2) \sim 0.05 - 0.06$.
The latter is roughly the $O$-$\phi$
mixing angle  $\sin\theta_{O\phi}$.
We note that $\theta_{O\phi} \simeq 3^\circ - 3.4^\circ$
is very close to the deviation from
ideal $\omega$-$\phi$ mixing, $\delta\cong 3.7^\circ$.
Thus, OZI violation in $1^-$ nonet is probably rooted in the heaviness of $O$!

It is fascinating to mention another recent OZI violating ``anomaly".
%in $\bar p p \to \phi + X$ vs. $\omega + X$ final states.
Several experiments have studied $\bar p p \to \phi + X$ vs. $\omega + X$
with $\bar p p$ annihilating at rest \cite{Ellis}.
One expects
\begin{equation}
R_X \equiv {\sigma (\bar p p \to \phi + X)
            \over \sigma (\bar p p \to \omega + X)}
         \sim \tan^2\delta \ltap 1\%, \nonumber
\end{equation}
which seems to be respected in most cases,
but with two prominent exceptions \cite{CBAR}:
\begin{equation}
R_\pi \simeq 0.1,\ \ \ \ R_\gamma \simeq 0.24.
\label{RXexp}
\end{equation}
These two cases proceed via specific initial states \cite{Ellis,CBAR},
$^1S_0(\bar pp)\to V\gamma$
and $^3S_1(\bar pp)\to V\pi$.
It is plausible that the common $I=0$, spin 1, excited $\bar pp$ system
annihilates completely into 3 gluons
without leaving behind some $q\bar q$'s.
We conjecture that  there is a substantial
resonance contribution (see Fig. 3) from
 $^1S_0(\bar pp) \to (\bar p p)^*_{I=0}\;
[ \to O \to\phi,\ \omega ] + \gamma$
and
$^3S_1(\bar pp) \to (\bar p p)^*_{I=0}\;
[ \to O \to \phi,\ \omega ] + \pi$,
where $O$ dominance gives the $SU(3)$ prediction
\begin{equation}
R_X = (1/\sqrt{2})^2 = 1/2.
\label{RXO}
\end{equation}
The experimental results for $R_\pi$ and $R_\gamma$, eq. (\ref{RXexp}),
should necessarily be smaller since $f(m_\omega^2) > f(m_\phi^2)$,
and since there should be more channels
(e.g.  ``incomplete" $\bar pp$ annihilation)
for $\omega$ final states,
especially for $\omega\pi$.
As for other modes $X = \eta,\ \rho,\ \omega, \pi\pi$, etc.,
they typically involve more partial waves and
there is considerably more cross section for
final states without $s\bar s$.
Our explanation of the anomalously large cross section for
$S$-state $\bar p p \to \phi \gamma$ and $\phi\pi$,
though qualitative, is cogent and simple compared to
most other \cite{Ellis} model explanations:
$O$ mediates OZI violation.
Since $O\to p\bar p$ width is rather small,
the strategy may be to search for its $0^{-+}$ partner as
a bump \cite{AGK} in $p\bar p$ cross section around $\sqrt{s} \simeq$ 3 GeV.

Let us summarize the main points of this paper.
The vector glueball state $O$, postulated 21 years ago, is alive and well.
As the $0^{++}$ glueball $G$ seems to have emerged
with $m_G \sim 1500-1700$ GeV, we argue that
$m_O$ is plausibly above 3 GeV.
The persistent absence of 
$\psi^\prime \to \rho\pi,\ K^*\bar K$ modes compared with
$B(J/\psi \to \rho\pi,\ K^*\bar K)\sim 1\%$
(the ``$\rho\pi$ anomaly") then strongly suggest the mass range for $O$
as given in eq. (\ref{Orange}).
Absence of distortion in
the recent BES energy scan of $J/\psi \to \rho\pi$
does {\it not} rule out $m_O\simeq m_{J/\psi}$,
but serves to put a lower bound on $\Gamma_O$,
while consistency requires 
$B(O\to \rho\pi) \gtap \mbox{few}\times B(J/\psi \to \rho\pi)$.
The range for $\Gamma_O$ is also summarized in eq. (\ref{Orange}).
A consistent decay picture for $O$ emerges,
where few $\% \ltap B(O\to \rho\pi) \ltap 25\%$ is a dominant but not predominant mode.
The $O\to e^+e^-$ mode is very suppressed, at the eV level,
while $K^+K^-$ and $p\bar p$ modes are quite suppressed
compared to $VP$ modes.
The prominence of $J/\psi \to VP$, $VT$
and $\eta_c\to VV$ modes is explained
by resonance enhancement due to nearby
$1^{--}(0)$, $1^{--}(2)$ and $0^{-+}(1)$ 3-$g$ glueball states
(number in parenthesis is $gg$ pair spin)
predicted by potential models.
The OZI suppression of $\phi\to \rho\pi$ is traced to the heaviness of $O$.
The $S$-state $\bar pp \to \phi\gamma$ and $\phi\pi$ annihilation anomaly
may be due to $O$ dominance,
which is facilitated by the limited number of channels.
The search for $O$ should continue
in $\psi^\prime \to \pi^+\pi^- + (\rho\pi$, $K^*\bar K)$,
perhaps via $\Upsilon(1S) \to G + O$,
and in $\bar p p$ annihilation in flight.
Once clearly seen, the glueball nature of $O$ should be unequivocal.

\acknowledgments
This work is supported in part by grant NSC 86-2112-M-002-019
of the Republic of China.
We thank J. M. Cornwall and A. Soni for initiating us on this path a long time ago,
and S. J. Brodsky, J. M. Izen, J. Li, S. L. Olsen, S. R. Sharpe, W. H. Toki,
P. Wang, and especially G. D. Chao, Y. F. Gu and S. F. Tuan 
for numerous discussions and communications.

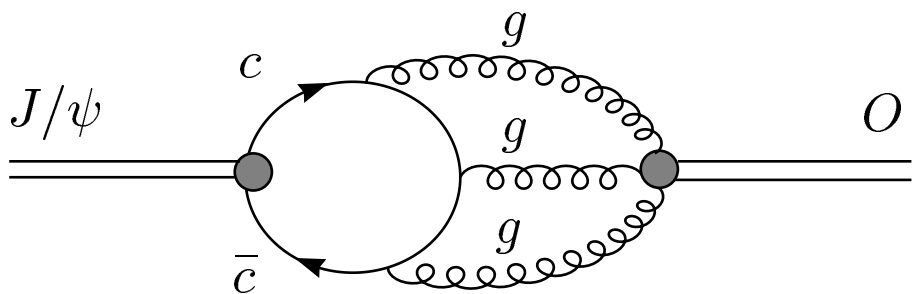
\begin{figure}
\caption{Mechanism for $J/\psi$-$O$ mixing. Likewise for other $V$-$O$ mixings.}
\label{fig1}
\end{figure}

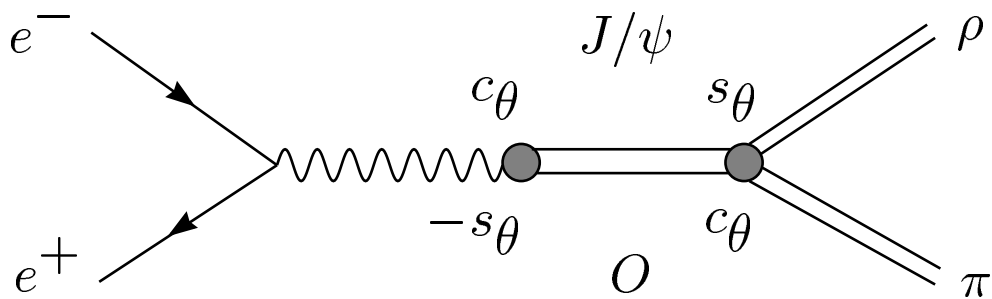
\begin{figure}
\caption{$e^+e^-\to\rho\pi$ via $J/\psi$ and $O$ intermediate states.}
\label{fig2}
\end{figure}

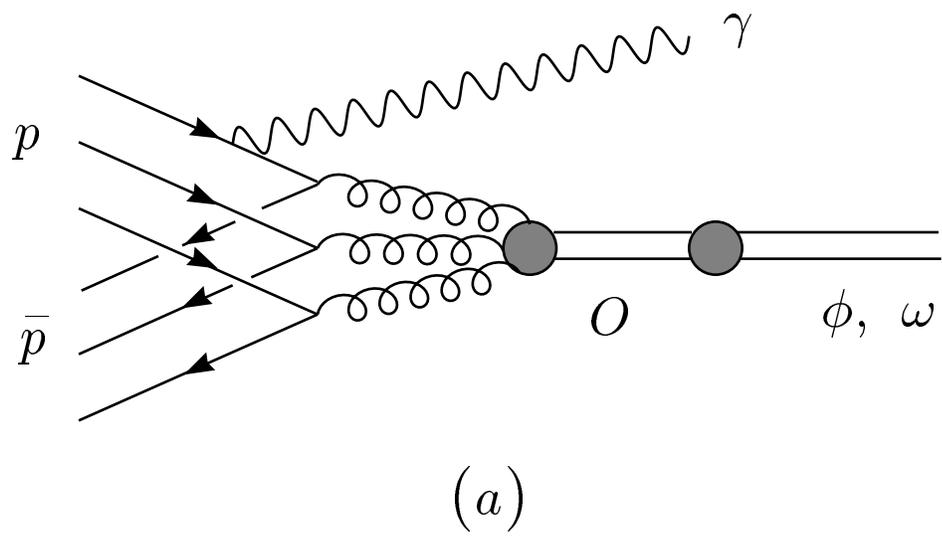
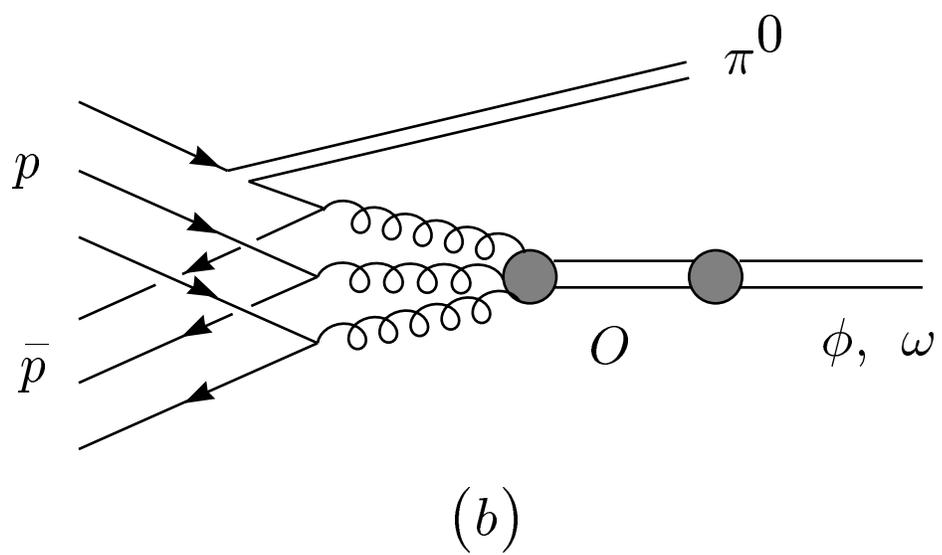
\begin{figure}
\caption{Scenario for $S$-state $\bar pp\to V \gamma,\ V \pi^0$ via $O$ dominance.
}
\label{fig3}
\end{figure}


\begin{references} 
%
\bibitem{Landua} R. Landua, ``{\it New Results in Spectroscopy}", plenary talk
at the International Conference on High Energy Physics (ICHEP96),
Warsaw, Poland, July 1996; and references therein.
%
\bibitem{Zhu} The spin of $f_J(1710)$ is still controversial.
See, for example, J. Z. Bai et al. (BES Collaboration),
to appear in Phys. Rev. Lett.;
I thank Y. C. Zhu for an advance copy.
%
\bibitem{UKQCD} %F.E. Close et al., preprint RAL-96-040;
G. S. Bali et al., Phys. Lett. {\bf B 309}, 378 (1993).
%
\bibitem{Weingarten} H. Chen et al.,
Nucl. Phys. (Proc. Suppl.) {\bf B34}, 357 (1994).
%
\bibitem{AC} C. Amsler and F .E. Close,
Phys. Lett. {\bf  B},  (1995); Phys. Rev. D {\bf 53}, 295 (1996).
%
\bibitem{SVW} J. Sexton, A. Vaccarino and D. Weingarten,
Phys. Rev. Lett. {\bf 75}, 4563 (1995).
%
\bibitem{xi} J. Z. Bai et al. (BES Collaboration), Phys. Rev. Lett.
{\bf 76}, 3502 (1996);
J. Li, talk presented at ICHEP96.
%
\bibitem{Farrar} M. B. \c{C}akir and G. R. Farrar,
 Phys. Rev. D {\bf 50}, 3268 (1994).
%
\bibitem{HS} W. S. Hou and A. Soni,  Phys. Rev. Lett. {\bf 50}, 569 (1983);
Phys. Rev. D {\bf 29}, 101 (1984).
%
\bibitem{Franklin} M. Franklin et al. (MARK II Collaboration),
Phys. Rev. Lett. {\bf 51}, 963 (1983).
%
\bibitem{AP} The BES Collaboration, ``Charmonium decays to axialvector plus pseudoscalar mesons", contributed paper to ICHEP96.
%
\bibitem{Gu1} Y. F. Gu, talk presented at the American Physical Society
Division of Particles and Fields 1996 Meeting (DPF96),
Minneapolis, USA, August 1996.
%
\bibitem{scan} J. Z. Bai et al. (BES Collaboration), Phys. Rev. D
{\bf 54}, 1221 (1996).
%
\bibitem{BLT} S. J. Brodsky, G. P. Lepage and S. F. Tuan,
Phys. Rev. Lett. {\bf 59}, 621 (1987).
%
\bibitem{FN} P. G. O. Freund and Y. Nambu,
Phys. Rev. Lett. {\bf 34}, 1645 (1975).
%
\bibitem{OZI} S. Okubo, Phys. Lett. {\bf 5}, 165 (1963);
G. Zweig, CERN Report No. 8419/TH412 (1964);
I. Iizuka, Prog. Theor. Phys. Suppl. 37, 21 (1966).
%
\bibitem{Cornwall} J. M. Cornwall, Phys. Rev. D {\bf 26}, 1453 (1982).
%
\bibitem{comm2} If the limits of eq. (\ref{psip}) continue to drop,
perhaps one should check whether
$f_{O\psi^\prime}/f_{O\psi}$ is also suppressed \cite{HS},
i.e. the overlaps $\langle \psi^\prime\vert O \rangle$ and
$\langle \psi\vert O \rangle$ should be calculated.
%
\bibitem{CS} J. M. Cornwall and A. Soni,  Phys. Lett. {\bf 120 B}, 431 (1983).
%
\bibitem{mO} To conform with eq. (\ref{BLT}),
the {\it premixed} $0^{++}$ glueball should lie between $1450 - 1530$ MeV.
Hence, we prefer $f_0(1500)$ over $f_0(1720)$ as the scalar glueball.
However, uncertainties in the potential model should
be no less than the lattice case.
%
\bibitem{AGK} M. Anselmino, M. Genovese and D. E. Kharzeev,
Phys. Rev. D {\bf 50}, 595 (1994).
The two gluon $0^{-+}$ state,
which has mass of order \cite{Landua} 2200 MeV,
is not considered.
%
\bibitem{CGT} K. T. Chao, Y. F. Gu and S. F. Tuan,
preprint BIHEP-TH-93-45, December 1993 (unpublished); 
Comm. Theor. Phys. {\bf 25} (1996) 471.
%
\bibitem{radtail} The radiative tail could
more easily hide an $O$ state above $m_{J/\psi}$.
See Fig. 2 of ref. \cite{scan}
%
\bibitem{update} A reanalysis is underway (Y. F. Gu, private communication),
and we await the results.
%
\bibitem{Olsen} S. L. Olsen, talk presented at Quarkonium 1996 meeting held at
the University of Illinois at Chicago, June 15, 1996, to appear in
Int. J. Mod. Phys. {\bf A}.
%
\bibitem{BL} S. J. Brodsky and G. P. Lepage,
Phys. Rev. D {\bf 24}, 2848 (1981).
%
\bibitem{Ellis} For discussion on data, as well as for references on
other theoretical explanations, see
J. Ellis et al.,
Phys. Lett. {\bf B 353}, 319 (1995).
%
\bibitem{CBAR} C. Amsler et al. (Crystal Barrel Collaboration),
Phys. Lett. {\bf B 346}, 363 (1995).
%
\end{references}
\end{document}